\documentclass[article]{elsarticle}

\usepackage{lineno,hyperref}
 \pdfoutput=1

\journal{Chinese Journal of Physics}

\begin{document}

\begin{frontmatter}

\title{New level of precision in study of neutron beta-decay. \\
Theoretical basics and experimental backgrounds
}

\author{V.V. Vasiliev \fnref{myfootnote}}
\address{NRC "Kurchatov Institute" - ITEP\\
25, B. Cheremushkinskaya, Moscow, Russian Federation}
\fntext[myfootnote]{e-mail basil\_v@itep.ru}

\begin{abstract}
The article provides a theoretical substantiation for a significant increase in the level of accuracy in determining the neutron lifetime using an alternative concept of neutron beta decay. Neutrons are distributed among different subsets depending on the sign of the scalar product of the neutron spin vector and the momentum of the electron emitted during decay. Therefore, the neutron lifetime can be  determined separately for each of the subsets as the inverse frequency of its decay, which is to be averaged over the number of neutrons in this subset. The total neutron lifetime on the total set of neutrons is calculated by averaging the partial (basic) lifetimes, considering their weights. The weights of the basic lifetimes are calculated by two weighing methods, leading, respectively, to the so-called lifetime of "nonpolarized" neutrons, the weighted average lifetime of neutrons and the central lifetime, i.e. the arithmetic mean for the basic lifetimes of neutrons. The dependence between different average neutron lifetimes through the integral asymmetry parameter calculated by using known experimental data leads to simple analytical expressions. The numerical estimate of the weighted average neutron lifetime gives a value that is in good agreement with the results of the well-known experiments, which proves the validity of the proposed concept of neutron beta decay. This article includes necessary conditions for a new experiment to bring the neutron lifetime determination to a new level of accuracy.

\end{abstract}

\begin{keyword}
\texttt{neutron}\sep decay\sep electron \sep asymmetry\sep  set theory
\end{keyword}

\end{frontmatter}

\section*{Introduction}
\label{intro}
One of the mysteries of modern physics is the inexplicably wide variety of experimental data on the measurement of the neutron lifetime. The case is that the measured values \cite{Zyla2020-1} fill two intervals: 875 - 900 s and 900 - 937 s. Such a variety of values obtained on neutron beams from different reactors in different countries by teams of qualified researchers requires careful analysis and explanation. 

An essential feature of the experiments of the last thirty years is a refusal to use registration of electrons when measuring the neutron lifetime in beta decay. The main emphasis was then placed on recoil protons in decay, and then experiments using storage of ultracold neutrons before beta decay in traps of various types took the lead. A possible reason for such a transition was an explicit or implicit desire to save the experiment from the uncertainty that the electron-spin asymmetry of neutron decay could introduce into the determination of the lifetime when registering electrons. In the experiments, proton traps were used, in which recoil protons from neutron decay were accumulated before being thrown onto the detector and then traps of ultracold neutrons where exactly the number of neutrons was recorded depending on the storage time became  main instruments. These methods of accumulation led to a certain way of forming an average value for the neutron lifetime, integrating in different ways all the details of the beta decay, in particular the asymmetry of electron emission. In this work, the goal is to return the study of the neutron beta decay to the registration of decay electrons in order to reveal all the details of the process and, due to these details, significantly increase the accuracy of determining the lifetime and, possibly, find an explanation for the above range of values.

The problem is that in order to improve the accuracy of the neutron lifetime, both from the point of view on its physical essence as well as in measuring it in experiments, it is necessary to adapt the basic known relations, describing the asymmetry of electrons emitted by neutrons, to the experimental realities. 

Improving accuracy becomes achievable due to the introduction of more accurate ideas about the neutron beta-decay with the involvement of the basic representations of the set theory and the theory of probability. It is important to mention that the possible application of new methods for recording the electron counting rates will make it possible to prevent distortion or artificial simplification of the decay pattern at the stage of measurements. Based on the literature data, numerical estimates prove the validity of introduced refinements.

\section {Theory of two-channel neutron decay. Consequences for the neutron lifetime} 
\label{sect1:THEORY}
The generally accepted description of the probability of a neutron beta decay is the Jackson-Treiman-Wilde formula (1957). In a simplified form it is usually represented by the dependence of the probability $W$ on the angle $ \theta_e $ between the direction of electron emission and the direction of the neutron spin \cite{Jacks1957}:
$$
dW(E_e,\Theta_e)= dE_e \cdot d \Theta_e \cdot W_0 \left(1+A\cdot \frac{v_e}{c}
\cdot \cos \theta_e \right) \mbox{,} \eqno (1)
$$
where $E_e $ is the electron kinetic energy, $A$ is the correlation coefficient of the electron emission with the direction of the neutron spin, $\theta_e$ is an electron emission angle relative to the direction of the neutron spin, 
$\Theta_e$ is a solid angle of electron emission, $\frac{v_e}{c}$ is the speed of the electron relative to the light speed, and $W_0$ is a constant.
The value of the coefficient $A$ is known from many experiments. Below we used two values of $A$: the value from 2010 year $A_{2010} = - 0.1173 \pm 0.0013$ \cite{PDG2010} and the value from 2020 year $A_{2020} = - 0.11958 \pm 0.00012$ \cite{Zyla2020-2}.

For each neutron, formula (1) ascribes the possibility of decay with an arbitrary direction of the electron momentum relative to the neutron spin. The probability of neutron decay with an electron escaping against the direction of the neutron spin is higher than with the electron momentum in the direction of the spin. Thus, formula (1) shows that the number of neutrons, which decay with electrons against the direction of the spin, is more than the number of those that decay with the emission of electrons in the direction of the neutron spin. This means that as a result of the decay of neutrons, two different final states are formed, differing in the sign of the scalar product $\left(\vec s_n \cdot \vec p_e\right)$, where the vector $\vec s_n$ is the spin of the neutron and the vector $\vec p_e$ is the momentum of the electron. Hence, it is legitimate to consider the beta decay of a neutron as a two-channel process. It is exactly this interpretation of beta decay that is discussed in this work. 

Depending on the final state during decay, we distinguish two types of neutrons. Neutrons of the first type - the so-called $L\mbox{–neutrons}$ - are able to emit electrons at the moment of decay against the direction of their own spin. In contrast to them, neutrons of the second type, $R \mbox{–neutrons}$, are able to emit electrons in the direction of their own spin in time of decay.

Obviously, these two types of neutrons have different properties just at the moment of their decay, which makes it possible to distinguish two implicit subsets corresponding to two types of neutrons in the total set of neutrons. Then, any set of neutrons $T$ with the number of neutrons $N_T$ can be considered as the sum of two subsets of neutrons: the subset of $L\mbox{–neutrons}$ with the number of neutrons $N_L$ and the subset of $R \mbox{–neutrons}$ with the number of neutrons $N_R$, moreover $N_T=N_L+N_R$. The value of the ratio $\frac{\bar v_e}{c}$ , where $\bar v_e$ -is the electron velocity averaged over the electron spectrum, is assumed to be equal for decays in both subsets. Let us point out that the ratio $\frac{\bar v_e}{c}$- is the eigenvalue of the helicity of electrons, i.e. $\frac{\bar v_e}{c}$ - is the value of the average projection of the electron spin on the electron momentum. Another difference from the concept of a generalized particle, which is the base for formula (1), is the separate determination of the decay constants on the indicated subsets of neutrons. On the subset of $L\mbox{–neutrons} $ with the number of neutrons $N_L$, the decay constant is defined as the reduced counting rate on this particular set: 
$$
\lambda_{nL}=\frac{1}{N_L}\cdot\frac{dN_L}{dt}\mbox{.} \eqno (2.1)
$$
Similarly, on the subset of $R \mbox{–neutrons}$ with the number of neutrons $N_R$, the decay constant is 
$$
\lambda_{nR}=\frac{1}{N_R}\cdot\frac{dN_R}{dt}\mbox{.} \eqno (2.2)
$$
The total decay constant $\lambda_{nT}$ is defined here on the total set $T$ as the count rate reduced to the total number of neutrons of this set:
$$
\lambda_{nT}=\frac{1}{N_T}\cdot\frac{dN_T}{dt}\mbox{.} \eqno (2.3)
$$
Then, after differentiating the sum $N_T=N_L+N_R$ , the following relation holds:
$$
\lambda_{nT}=\lambda_{nL} \cdot \frac{ N_L }{N_L+N_R}+\lambda_{nR}\cdot \frac{N_R}{N_L+N_R}\mbox{.} \eqno (3.1)
$$
Comparison (3.1) with the full width rule \cite{Blatt}, \cite{Gold}, which for the total decay constant in parallel decay in the conventional form is written as the sum of partial constants:
$$
\lambda_{nT}=\sum_{i}\lambda’_{i} \mbox{,} \eqno (3.2)
$$
where $i$ - is the channel number, shows the following.
Partial constants in case (3.2) are determined with loss of generality, since 
$$
\lambda’_{i}=\frac{1}{N_T}\cdot \frac{dN_i}{dt}\mbox{,}
$$
so they contain different quantities in the denominator and under the sign of the derivative, while in the record of the full constant on the right side of (2.3) there is the same quantity, both in the numerator and in the denominator, i.e. the total number of particles $ \sum_{i}^k N_i$.
Thus, it was shown in (3.1) that the terms in the sum (3.2) contain hidden parameters which will be revealed if the partial constants are determined correctly in accordance with the definition of the reduced decay frequency, i.e.
$$
\lambda_i=\frac{1}{N_i}\cdot\frac{dN_i}{dt}\mbox{.}
$$
These hidden parameters are weights like those used in the record (3.1), and after correction the sum (3.2) looks like this 
$$
\lambda_{nT}=\sum_{i=1}^k \lambda_i \cdot \frac{N_i}{\sum_{i=1}^k N_i}=\sum_{i=1}^k\lambda_i \cdot \omega_i \mbox{,}
$$
where the weight $\omega_i$ for the $i\mbox{-th}$ channel is 
$$
\omega_i=\frac{N_i}{\sum_{i=1}^k N_i}\mbox{.}
$$

A weight is a probability of any particle to belong to the subset of particles that have the property of decay into a given channel with a rate reduced to the number of particles belonging only to this subset.
It should be noted once again that in the definitions (2.1) and (2.2) and in the above correction of the full width rule under the sign of the derivative and in the denominator is the same value - the number of neutrons with the ability to decay into the given channel.

The definition of the partial (basic) decay constant as the reduced decay rate of particles with ability to decay into the given channel is a new refinement of the decay constant concept for the case of parallel channels.

Since (3.1) is the formula for the weighted average, the notation $\lambda_{nW}\equiv \lambda_{nT}$
will be used below. Then (3.1) takes the form
$$
\lambda_{nW}=\lambda_{nL} \cdot \omega_L+\lambda_{nR} \cdot \omega_R\mbox{.}\eqno (4)
$$
A decay channel weight is the ratio of a partial constant for a given channel to the sum of partial constants for all channels \cite{Blatt}, \cite{Gold}. Then, in the notation (3.1), one can use the relations 
$$
\frac{N_L}{N_L+N_R}=\frac{\lambda_{nL}}{\lambda_{nL}+\lambda_{nR}}
$$
and 
$$\frac{N_R}{N_L+N_R}=\frac{\lambda_{nR}}{\lambda_{nL}+\lambda{nR}} \mbox{.}
$$
According to this definition the weighted average neutron decay constant (4) takes the form:
$$
\lambda_{nW}=\frac{\lambda^2_{nL}}{\lambda_{nL}+\lambda_{nR}}+
\frac{\lambda^2_{nR}}{\lambda_{nL}+\lambda_{nR}}\mbox{.}
$$
It is convenient to represent decay constants (2.1) and (2.2), using formula (1), through the quantities 
$$
\lambda_{n0}=\frac{\lambda_{nL}+\lambda_{nR}}{2}
$$ and
$$
\Delta=|A| \cdot \frac{\bar v_e}{c}\mbox{,}
$$ 
which, taking into account the average angles of electron emission for $L \mbox{-neutrons}$ and $R \mbox{-neutrons}$, gives 
$$
\lambda_{nL}=\lambda_{n0}\cdot (1+\Delta) 
$$ 
and 
$$
\lambda_{nR}=\lambda_{n0}\cdot (1-\Delta) \mbox{.}
$$
The lifetime $\tau_{n0}$ is
$$
\tau_{n0}=\frac{1} {\lambda_{n0}}
$$ 
and will be conventionally called as the lifetime of "nonpolarized" neutrons. Partial (basic) lifetimes for subsets of $L \mbox{-neutrons}$ and $R\mbox{-neutrons}$ are defined in the usual way $\tau_S=1/ \lambda_S $, where $S=nL, nR$. The lifetimes of $L\mbox{-neutrons}$ and $R\mbox{-neutrons}$ take the form
$$
\tau_{nL}=\tau_{nCentre}\cdot (1-\Delta) \mbox{,}\eqno(5.1)
$$
$$
\tau_{nR}=\tau_{nCentre}\cdot (1+\Delta) \mbox{,}\eqno(5.2)
$$
where $ \tau_{nCentre}=\tau_{n0}/(1-\Delta^2)$.
For consistency, it is convenient to express $\tau_{n0}$ through $\tau_{nCentre},\Delta$:
$$
\tau_{n0}=\tau_{nCentre}\cdot \left(1-\Delta^2\right)\mbox{.}\eqno(6)
$$

The lifetime of "nonpolarized" neutrons is also expressed as a weighted average of the basic lifetimes 
$\tau_{nL}$ and $\tau_{nR}$ like (4) by the formula 
$$
\tau_{n0}=\tau_{nL}\cdot \omega_L+\tau_{nR}\cdot \omega_R \mbox{,} \eqno(7)
$$
where the weights indicated in (7) and introduced earlier in (4) get the form
$$
\omega_L=\frac{\tau_{nR}}{\tau_{nL}+\tau_{nR}}=\frac{1}{2}+\frac{\Delta}{2}\mbox{,}\eqno(8.1)
$$
$$
\omega_R=\frac{\tau_{nL}}{\tau_{nL}+\tau_{nR}}=\frac{1}{2}-\frac{\Delta}{2}\mbox{.} \eqno(8.2)
$$
At the same time, the weighted average neutron lifetime is defined as $\tau_{nW}=~1/\lambda_{nW}$ and is expressed through the values $\tau_{nCentre},\Delta$ by the formula:
$$
\tau_{nW}=\tau_{nCentre}\cdot \left(1-\frac{2 \cdot \Delta^2}{1+\Delta^2}\right) \mbox{.} \eqno(9)
$$
For the weighted average neutron lifetime, the following expression is also valid:
$$
\tau_{nW}=\tau_{nL}\cdot W_{nL}+\tau_{nR}\cdot W_{nR} \mbox{,}\eqno(10)
$$
where the $\tau \mbox{-weight}$ of the $L \mbox{-channel}$ and the $\tau \mbox{-weight}$ of the $R\mbox{-channel} $ have the following dependence on $\Delta$:
$$
W_{nL}=\left(\frac{1}{2}+\frac{\Delta}{1+\Delta^2}\right) \mbox{,} \eqno(11.1)
$$
$$
W_{nR}=\left(\frac{1}{2}-\frac{\Delta}{1+\Delta^2}\right) \mbox{.} \eqno(11.2)
$$
So the weights (8.1), (8.2) for the lifetime of nonpolarized neutrons and the weights (11.1), (11.2) for the average weighted lifetime differ due to the definitions of the quantities $\lambda_{n0}$ and 
$\lambda_{nW}$. Expressions (5.1), (5.2) show the symmetry of the decay pattern around the centre - the point $\tau_{nCentre}$ on the axis of the trial lifetime. Then a functional describing the decay depending on the trial lifetime must be symmetric about the centre. If the functional is symmetric and, at the same time, has a minimum to the left of the centre of symmetry at the point of the weighted average neutron lifetime (9), then, by virtue of symmetry, it will also have a reflected minimum to the right of the centre at the point $\tau_{nW}^M$. This so-called mirror neutron lifetime $\tau_{nW}^M$ is expressed by the formula:
$$
\tau_{nW}^M=\tau_{nCentre} \left(1+\frac{2 \cdot \Delta^2}{1+\Delta^2} \right) \mbox{.} \eqno (12)
$$ 

The reasons similar to the above introduction of lifetime of "nonpolarized" neutrons lead to the hypothesis of the existence of one more parameter in the totality of the temporal characteristics of neutrons - the mirror lifetime for "nonpolarized" neutrons
$\tau_{n0}^M=~\tau_{nCentre}\cdot \left(1+\Delta^2\right)$.
This gives rise to the criteria for constructing the error functional required to describe the experimental data depending on the trial lifetime $\tau$:

1. Symmetry about the central point $\tau_{nCentre}$.

2. The minima at points $\tau_{nW}$ and $\tau_{nW}^M $ are symmetrical about the centre.

3. Correspondence of the error functional to the number of degrees of freedom.

Thus, the parameters describing the beta decay of neutrons, by means of simple calculations, are reduced to the elementary concepts of set theory and probability theory \cite{Kolm}. Therefore, the total decay constant in parallel decay is the weighted average value of the basic lifetimes in the set of channels. The detailed description of neutron decay as a parallel decay in the proposed alternative model leads to new physical quantities that can be estimated numerically on the base of literature data.

In addition, it will be shown below that in the selection of one of the channels in parallel decay and with a precise study of the counting rates of decay products in this channel \cite{VV1}, it is possible to measure the basic lifetime defined on the separate subset of particles.

\section {Estimation of the split effect in neutron decay} 
\label{sect2:EstSplit}
Thus, the effect of the splitting of the lifetime (5.1), (5.2) is associated with the asymmetry parameter $\Delta=|A|\cdot \frac{\bar v_e}{c}$. In this case, this parameter is the relative shift of the basic lifetimes. The second factor in this formula, the average velocity of electrons in units of the speed of light, is determined from the spectrum of electrons from neutron decay, which measured in \cite{Robsn}, where the spectrum of electrons was shown as symmetric in energy and, therefore, the maximum of the symmetric spectrum is located at the average value 
$ \bar T_e$ of the electron kinetic energy, $\bar T_e=391 \, \mbox{keV}$.
Using the formula,
$$
\left(\frac{\bar v_e}{c}\right)=\sqrt{\left(\frac{\bar T_e}{\bar T_e+m_e \cdot c^2}\right)
\cdot \left(2-\frac{\bar T_e}{\bar T_e+m_e \cdot c^2}\right)}
$$
where $m_e \cdot c^2=511 \, \mbox{keV}$, we obtain the estimate equals to 
$\left(\frac{\bar v_e}{c}\right)=0.824$.
For the electron-spin correlation coefficient A, two values, mentioned above $A_{2010}$ and $A_{2020}$ are used.  With the value $A_{2010} = -0.1173$, the square of the relative time shift is very close in size to ratio $1/107$, since  $\Delta^2 \approx 1/107.04$.
With the value $A_{2020} = -0.11958$, the square of the relative temporal shift is almost equal to the ratio $1/103$.  
For further calculations, the designation $\Delta^2=1/\eta$ is convenient. In the arguments $\tau_{nCentre},\eta$, formulas (9), (12) get simpler forms:
$$
\tau_{nW}\cdot \left(\eta+1\right)= \tau_{nCentre}\cdot \left(\eta-1\right)\mbox{,}\eqno(13)
$$
$$
\tau_{nW}^M\cdot \left(\eta+1\right)= \tau_{nCentre}\cdot \left(\eta+3\right)\mbox{.}\eqno(14)
$$

Considering the role of the parameter $\eta$ for the specified relationship between the neutron lifetimes, we will call it here as a split-parameter, i.e., the parameter characterizing the effect of splitting the neutron lifetime - the split effect. For the primary estimates of the above totality of neutron lifetimes, two values of the split-parameter are used:  $\eta=103$ and $\eta=107$.
Based on the values of the parameters $\tau_{nCentre}\mbox{,} \eta$, we can determine the basic lifetimes 
$\tau_{nL}\mbox{,}\tau_{nR}$ from the system of equations
$$
\left\{
\begin{array}{rcl}
\tau_{nL}+\tau_{nR}=2\cdot \tau_{nCentre} \mbox{,}\\
\tau_{nL}/\tau_{nR}=(\sqrt{\eta}-1)/(\sqrt{\eta}+1)\mbox{.}
\end{array} 
\right. \eqno(15)
$$
The numerical value for the central lifetime is given in the paper \cite{VV2}, which is unique in the sense that only there the experimental data on the electron counting rates are described by the error functional depending on the trial neutron lifetime. That article shows that this dependence is symmetric for the neutron decay and the center of symmetry coincides with the value $900.00\pm 0.15 \,\mbox{s}$. Considering the calculations of the previous section, we take this value here as an estimate of the central neutron lifetime $\tau_{nCentre}$.

Based on this result for the central lifetime and taking into account the two values of the split-parameter, the numerical estimates for values (6), (9), and (12) are as follows.

\begin{tabular}{|p{6cm} | c | c |}
\hline
\multicolumn{3}{| c |}{\textbf{Estimates of neutron lifetimes}}\\
\hline
Split-parameter, $\eta$ & 107 & 103 \\
\hline
Lifetime of “nonpolarized” neutrons, s & 891.59 & 891.26 \\
\hline
Weighted average neutron lifetime, s & 883.33 & 882.69 \\
\hline
Neutron mirror lifetime, s & 916.67 & 917.31 \\
\hline
\end{tabular}

Thus, when deriving the total decay constant from primary principles based on the elementary set theory and the probability theory \cite{Kolm}, a situation of contradiction with the concept of a generalized particle and the resulting full width rule arises. Considering here the neutron beta decay as a two-channel process is also a definite violation of the established traditions. However, two different final states in neutron beta decay are adopted by default after the theoretical discovery of electron-spin asymmetry in 1957 and received multiple experimental confirmations up to now. To prove the legitimacy of using the simple formulas obtained here when dividing the total set of neutrons into the two disjoint subsets, it is necessary to show that there is a fundamental possibility to determine experimentally the lifetimes on these different subsets. The next section is devoted to this problem.

\section {Determination of lifetime for a subset of neutrons} 
\label{sect3: Lifetime on subset }

The purpose of this section is to prove the fundamental possibility of determining the lifetime on a selected subset of neutrons from the counting rates of electrons.

Now we consider any facility or device as a neutron source, from an isotope neutron source to a reactor or an accelerator with a neutron-producing target. Let us install a vacuum chamber in a magnetic field with neutrons passing through. An electron detector monitors the chamber.  Electrons, which are the decay product of neutrons passing through this chamber or some area of the chamber, go to the electron detector along the magnetic field.

Let $N$ of neutrons is in the chamber volume in the time interval $dt$.
Neutrons passing through or being in a given area decay exponentially and their number changes as
$$
N(t)=N_0\cdot exp(-\lambda \cdot t) \mbox{,}\eqno(16)
$$
where $N_0$ is the initial number of neutrons at $t = 0$.
The quantity $\lambda$ is the decay constant and the reciprocal of the lifetime $\tau$, i.e.
$\lambda=1/\tau$.
In differential form, law (16) has the following expression,
$$
\frac{dN}{dt}=-\lambda\cdot N \mbox{.}\eqno(17)
$$
Where on the left is the rate of neutron loss in the monitored area: the minus sign on the right reflects the process of neutron loss. 
However, since the decay of one neutron is the production of one electron, the rate of electron production 
$\frac{dN_e}{dt}$ in this region is equal in magnitude to the rate of neutron loss and will be written with the plus sign on the right side as follows:
$$
\left|\frac{dN}{dt}\right|=+\lambda \cdot N \mbox{.}\eqno(18)
$$
In this case, electrons generated in the monitored area are collected on an electron detector with a certain efficiency, which we denote as $ \varepsilon_1$, and are registered by the detector with an efficiency of $ \varepsilon_2$. Let us describe the transport and registration of electrons by the detector, multiplying the left and right sides of (18) by the total efficiency $\varepsilon=\varepsilon_1\cdot \varepsilon_2$:

$$
\varepsilon \cdot \left|\frac{dN}{dt}\right|=\lambda \cdot \left(\varepsilon \cdot N \right)\mbox{.}\eqno(19)
$$
As a result, in equality (19), the electron-counting rate by the detector is on the left, and there is the product of the decay constant times by the value in parentheses on the right. The value in parentheses equals to the number of neutrons that the detector "sees", i.e., it receives signals from them in the form of registered electrons. It means there is the rule for detectors -"You only get from what you can see''. Let us repeat (19) in a simpler form, introducing the designation for the electron-counting rate $R_D$ and the number $N_D$ of neutrons, visible by the detector:$$
R_D=\lambda \cdot N_D \mbox{.} \eqno(20)
$$

Based on equation (20), we set the problem to determine the decay constant by measuring only the electron counting rates and completely rejecting any measurements of the number of neutrons in the right-hand side of (20), i.e. excluding measuring the number of neutrons, "visible" by the electron detector. To solve this problem, let us add a condition of varying the number of neutrons to ensure changes in the electron-counting rate in the wide range, the upper limit of which is several times higher than the lower limit. For the case of variations, equation (20) transforms into a system of equations. Omitting the index $D$ to simplify the notation and adding the error $\sigma_i$ in measuring the count rates in all series at each step and at all steps of variations this system receives the following form:
$$
\left\{
\begin{array}{rcl}
R_1 \pm \sigma_1& \approx &\lambda \cdot N_1\mbox{,}\\
R_2 \pm \sigma_2& \approx &\lambda \cdot N_2\mbox{,}\\
\ldots& \approx&\ldots \mbox{,} \\
R_k \pm \sigma_k& \approx &\lambda \cdot N_k\mbox{.}
\end{array} 
\right. \eqno(21)
$$
The initial number of neutrons changes in steps, therefore, the neutron numbers in the field of view of the detector also change in steps, passing in the process of variations successively $k$ values of $N_i$. In this case, only the electron-counting rates are measured and recorded in a continuous mode. At each stage of the number of neutrons, a sufficient number of independent series of measurements are carried out at low counting rates, i.e., counting small numbers of pulses in long reading intervals. The result of each independent series is a pair of numbers: the average count rate and the error in the each series. To ensure the accuracy of the count rate values, a high-frequency timer based on a stable frequency generator measures the readout time intervals.
Let us write a system of equations (21) for the set of $k$ neutron numbers, representing the numbers as members of an arithmetic progression $N_i=~\frac{1}{\mu} \cdot ~m_i$.  Here $\frac{1}{\mu}$ is the difference of the arithmetic progression or the step of a scale of neutron numbers, a decimal number; the integer number $m_i$ is the $i\mbox{-th}$ number of the scale division corresponding to the neutron number. 
We get the system of decay equations:
$$
\tau \cdot \left( R_i \pm \sigma_i \right) \approx \frac{1}{\mu} \cdot m_i \mbox{.} \eqno(22) 
$$
The step size of the scale $1/ \mu$ uniquely determines the set of integers $m_i$ - the numbers of the scale divisions corresponding to the array of pairs of numbers for the given value $\tau$. Here $i = 1, 2, \ldots k $ , $k $- is the full number of measurements, including all repeated measurements of the count rate at each step of the neutron number. The task is reduced to the choice of the optimal scale factor $\mu$. It is necessary to choose such a step of the neutron number scale $1/ \mu$ , which forms the optimal arithmetic progression, the selection of whose members provides the best description of the array of pairs $R_i,\sigma_i, i=1,2, \ldots k$. 
To estimate the sequence (scale) of neutron numbers, realized in a variation experiment, a scale adjustment operator has been constructed, which has the following form:
$$
\aleph_i =round \left[\frac{round \left[\mu \cdot \tau \cdot R_i,0 \right]}{\mu}, p \right] \mbox{.} \eqno (23)
$$
The operator $round [C, p ] $ is used to return the value of $C$, rounded to the $p\mbox{-th}$ decimal place. For example, $round[2.178, 2] = 2.18$; $round[2.1859, 3] = 2.186$; $round [2.19, 0] = 2$. The structure of the operator (23) exactly corresponds to the equation (22). Operator (23) forms a scale with a step of $1/\mu$. Using the trial lifetime $\tau$ and the scale factor $\mu$, the number of divisions corresponding to the count rate $R_i$ is calculated by the internal operator $round [\mu \cdot \tau \cdot R_i,0]$ by rounding to the nearest integer. In this case, the estimate $\aleph_i$ for the neutron numbers $N_i$ is made with a fixed approximation accuracy $p$, which fact is important to confirm the convergence depending on the accuracy at the solution point. 
The following error functional $F_{\mu,p}(\tau)$ is constructed for the required range of $\tau$ for different $p$:
$$
F_{\mu,p}(\tau)=\sum_{i=1}^k \frac{\left(R_i-\frac{1}{\tau} \cdot
\aleph_i(p,\mu,\tau) \right)^2}{\sigma_i^2} \mbox{.} \eqno (24)
$$
The functional  (24) is investigated for different $\mu$-factors, and the solution is an optimal value of the $\mu$-factor, which ensures the reduction of the functional at the minimum to unity, together with the coordinate of the minimum on the axis of the trial lifetime. 

Thus, the system of equations (22) is solved by the least squares method with one parameter, which is the difference of the arithmetic progression optimal for minimizing the error functional (24), which describes the experimental data array using the adjustment operator (23) for the decay scale tuning. 

The method was first proposed in \cite{VV3} and is applicable to any radioactive decay. Under the condition of the presence of a single channel or selection of one decay channel and with dense filling of a sufficiently wide range of counting rates of decay products (in this case, electrons), the integer approximation of neutron numbers is sufficient for solving, i.e., for $\mu=1$ or for any $\mu$ and $p = 0$. The only solution in this case coincides with the minimum of the parabolic dependence of the functional on the trial lifetime. Then, to determine the error of the solution, it is sufficient to determine the value of the $\mu\mbox{-factor}$ at which the minimum of the functional will be fixed on a level close to $\chi^2=1$ for any accuracy of the solution. In this case, the error is determined by the half-width of the parabola at a level determined by the number of measurements. 

However, as shown above, in the neutron case, the decay picture is more complex. Difference in the decay of $L\mbox{-neutrons}$ and $R\mbox{-neutrons}$ leads to the need to separate the electron counting rate from $L \mbox{-neutrons}$ in the form of an equation
$$
R_L\pm\sigma_L=\lambda_L \cdot N_L \mbox{,} \eqno(25.1)
$$
and the electron counting rate from $R \mbox{-neutrons}$ in the form
$$
R_R\pm\sigma_R=\lambda_R \cdot N_R \mbox{.} \eqno(25.2)
$$
The sum of the counting rates (25.1) and (25.2) is expressed in terms of the total number of neutrons with a coefficient in the form of a weighted average decay constant $\lambda_W$:
$$
R_L+R_R=\lambda_W \cdot \left(N_L+N_R \right) \mbox{.}\eqno(26)
$$
With an adequate method of averaging primary counts in the distribution of measured count rates, it is possible to save information on the decays of $L\mbox{-neutrons}$ (25.1) and $R\mbox{-neutrons}$ (25.2). As result of summation and averaging, the average value of the neutron lifetime described by the weighted average decay constant (26) is inevitably present. 

Therefore, the weighted average decay constant describes the change in the total number of neutrons in decay. Calculation of the dependence of functional (24) on the trial lifetime for different values of the scale step, considering the above guidelines, solves the problem of minimizing the functional, finding the center of its symmetry, and determining the value of the weighted average neutron lifetime. 

In the case of using a polarized neutron beam, the scale adjustment method can be applied separately when registering electrons emitted against the direction of neutron polarization (25.1) and registering electrons emitted in the direction of neutron polarization (25.2). In this case, a strict spatial separation of the decays of $L\mbox{-neutrons}$ and $R\mbox{-neutrons}$ is realized, which makes it possible to determine their lifetimes separately by the above method. The control experiment is reduced to rearrangement of detectors or to spin-flip of the initial polarized beam by $180^{\circ}$ and repetition of measurements. To vary the initial number of neutrons in the case of a neutron beam, experiments can use the known optical methods that provide necessary changes of the neutron flux with the required fixation. In the case of a "white" neutron beam or the use of non-reactor sources, the method is also applicable if the method for measuring the electron-counting rate meets the requirements for sensitivity to a difference of decay frequencies and provides a differential measurement of decay frequencies in the realized range.

The values of the $ \mu $ -factor in the case of the neutron decay are determined by the estimate of the split-parameter from the previous section.

\section{Discussion of the results} 
\label{sect4:DISRT}
Now then, considering the neutron beta decay as an example of two-channel decay on the basis of set theory and probability theory, describes the asymmetry of beta decay in the concept of two disjoint subsets of neutrons and leads to the two basic lifetimes of neutrons.

The integral asymmetry coefficient  $\Delta$ plays the role of a relative shift of the basic lifetimes and allows to calculate the weights of two neutron decay channels. The total neutron lifetime is calculated as a weighted average of the indicated basic lifetimes. In addition to the weighted average, an important role is played by the arithmetic average of the basic lifetimes, the difference of which from the weighted average value is evidence of the existence of a lifetime split effect in the neutron beta decay. 

The value of the weighted average lifetime is in a good agreement with the results of the experiment of NIST (USA) $\tau_n=887.7\pm 1.2 (stat)\pm 1.9 (syst) \, \mbox{s} $ \cite{Yue2013}  and the experiment of PNPI (Russian Federation) $\tau_n=881.5 \pm 0.9 \quad \mbox{s}$ \cite{Sereb2018}. 

Let us explain here the combination of the words “in a good agreement”. The doubled total error of the result \cite{Yue2013} is 4.5 seconds. The doubled error of the result \cite{Sereb2018} is 1.8 seconds. The lower confidence interval (at $2\sigma \mbox{-level}$) of the first result coincides with the upper limit of the second result within 0.1 second. The result for $A_{2010}$ lies precisely at this intersection of the boundaries of the confidence intervals, i.e., it is separated from both results by exactly two errors of each of them.  The result from $A_{2020}$ is less only by 0.7 s. By the way, this coincidence is a proof of good mutual agreement between the indicated results. 

Thus, the concept of two-channel neutron decay put forward in this work based on set theory and a more accurate determination of the main decay parameters is confirmed by independent experiments. With an adequate definition of the neutron decay constants as characteristics of two sets of neutrons with different final states the application of the weighted average value, what is usual for the theory of probability, works successfully for the total neutron lifetime. 

It is possible to extrapolate this approach to other multichannel processes in order to refine the values of important nuclear physics constants. 

According to the proposed concept, new physical quantities are introduced to describe the beta decay of a neutron - the basic lifetimes of neutrons - the lifetime of $L \mbox{-neutrons}$ emitting an electron during decay against the direction of the neutron spin and the lifetime of $R \mbox{-neutrons}$ emitting an electron in the direction of the neutron spin during decay. 

It is shown that the basic lifetimes of neutron can be conveniently expressed in terms of the central neutron lifetime and the split parameter $\eta=1/\Delta^2$, $\eta=107$ or $\eta=103$,
The lifetime of $L\mbox{-neutrons}$ in these parameters is 
$$
\tau_{nL}=\tau_{nCentre}\cdot \left(1-\frac{1}{\sqrt{\eta}}\right) \mbox{.}
$$
The lifetime of $R\mbox{-neutrons}$ in these parameters is
$$
\tau_{nR}=\tau_{nCentre}\cdot \left(1+\frac{1}{\sqrt{\eta}}\right) \mbox{.}
$$ 
The total neutron lifetime is calculated here as the weighted average of the basic lifetimes $\tau_{nL}$ and $\tau_{nR}$, taken with the following weights 
$$
W_{nL}=\left(\frac{1}{2}+\frac{\sqrt{\eta}}{\eta+1}\right)
$$
and
$$
W_{nR}=\left(\frac{1}{2}-\frac{\sqrt{\eta}}{\eta+1}\right)\mbox{,}
$$
respectively.
The values of the basic lifetimes of the neutron are also determined by solving the system of equations (15) and equal to the following values in dependence of $ \eta$- value.

\begin{tabular}{|p{5cm} | c | c |}
\hline
\multicolumn{3}{| c |}{\textbf{Estimates of basic neutron lifetimes}}\\
\hline
Split-parameter, $\eta$ & 107 & 103 \\
\hline
Lifetime of L-neutrons, s & 812.99 & 811.32 \\
\hline
Lifetime of R-neutrons, s & 987.01& 988.68 \\
\hline
\end{tabular}

A completely new physical quantity is the mirror neutron lifetime $\tau_{nW}^M$. An indication of the existence of this physical constant is given here for the first time. However, an experimental result close to this value is known for a long time as equal to $918 \pm 14 \, \mbox{s}$ from the research \cite{Chr72}, where the neutron decay electron count was used. This result is traditionally placed in the PDG summary tables for the neutron lifetime in the lower part, which is not included in calculating the average value. The author, however, believes that this experimental result is physically justified.

The above estimate, obtained because of symmetry considerations about the central lifetime, can also be expressed by the following formula:
$$
\tau_{nW}^M=\tau_{nL}\cdot W_{nR}+\tau_{nR}\cdot W_{nL}\mbox{.}
$$

Nevertheless, the main result for comparison is the weighted average neutron lifetime, the estimate of which, as indicated above, is in full agreement with the known experimental data. 

It is impossible not to mention one more coincidence from the above triad of averages. In this work, for the first time, the problem of determining the so-called lifetime of "nonpolarized" neutrons is formulated, which is the inverse of the arithmetic mean of the decay constants of $L \mbox{-neutrons}$ and $R \mbox{-neutrons}$. This result coincides with the value of the neutron lifetime obtained by P.E. Spivak \cite{Spivak1988} in 1988: $\tau_n= 891\pm 9 \, \mbox{s}$. The value of the neutron lifetime error in that work coincides with the difference between the arithmetic mean neutron lifetime and the lifetime of the "nonpolarized" neutron. Moreover, this error interval at the lower limit coincides with the value of the weighted average neutron lifetime. Hence, the error corresponds to the structure of average values for the neutron lifetime, i.e., the error is physically substantiated, and the result of \cite{Spivak1988} generally also confirms the validity of the estimates obtained in the present paper. 

Discussing the two different lifetime results of the experiment of NIST, USA
$\tau_n=887.7\pm 1.2(stat)\pm 1.9 (syst) \, \mbox{s}$ \cite{Yue2013}  and the experiment of PNPI, Russian Federation  $\tau_n=881.5 \pm 0.9 \quad \mbox{s}$ \cite{Sereb2018} the author must give PDG2020 average value $879.4 \pm 0.6 \, \mbox{ s}$ to compare with. It disagreed at some level with the mentioned results. The reasons for this difference are quite explainable. The UCN-storage experiments always give lifetimes less than the true neutron lifetime. With increasing accuracy, these experiments give the lower approximation to the true lifetime, since containing side channels of neutron leakage. To prove the lack of side channels of leakage, you can only know the true lifetime with much higher accuracy. For example when neutrons are stored in magnetic traps, a bias in the estimate of the lifetime is inevitable due to the unaccounted systematic error, which is characteristic of the ”bowl”-type magnetic traps. This error is associated with the depolarization of ultracold neutrons at the nodes of an inhomogeneous magnetic field that keeps neutrons of only a certain polarization in the trap \cite {VladVV}. Depolarization occurs when a neutron passes through the field nodes, i.e. zero magnetic field, with a spin flip relative to the field. Spin-flip at the field nodes leads to the loss of neutrons i.e. reduces the storage time of neutrons in the trap. This effect is difficult to measure or calculate especially if it is only a few seconds versus the neutron lifetime of about 900 seconds. Mathematically, the presence of nodes in a”bowl”-type system consisting of a bottom and walls is very easy to prove, which was done in \cite{VVV}.  A detailed discussion of this source of a systematic error is beyond the scope of this work.

The results show that the neutron beta decay during the registration of decay products looks a little more difficult. However, the detailing of experimental study of this process by adequate methods leads to the level of accuracy of the order of hundredth seconds.  This level of accuracy will solve many contradictions and explain all the accumulated information about the beta decay of neutrons.

\section{Conclusion} 
\label{sect5:CONCL}
It is shown that the neutron beta decay is a unique example of two-channel process. A new element was the use of elementary set theory to describe the well-known effect of electron-spin asymmetry in neutron decay. The concept of a partial decay constant and a partial lifetime are refined as characteristics of a subset of particles that have the property of decay into a given channel. It is shown that the total neutron lifetime is a weighted average of the partial lifetimes, which leads to a more accurate formulation of the full width rule. Analytical expressions are obtained that relate the basic neutron lifetimes and the weighted average value of the neutron lifetime with the integral parameter of the neutron beta decay asymmetry.
Numerical estimates of the weighted average value of the neutron lifetime are obtained on the basis of the developed concept and known experimental data. Convincing agreement with the known independent experimental results was noted, which is proof of the validity of the concept of two lifetimes in the neutron beta decay.
The dependence of the key parameters of the neutron beta decay on the split~-parameter and the central lifetime are obtained. The ways of transition to a new level of precision are shown and substantiated due to the detailed study of the neutron beta decay. Theoretical estimates are given for the expected values for the two basic lifetimes of neutrons. 
The estimates agree with the preliminary experimental results \cite{VV4}.

\section{Acknowledgements} 
In this work, the author used the results obtained during the processing of experimental data 
on neutron beams from the ITEP heavy water reactor, and developed some approaches that 
were the subject of discussions in the ITEP laboratory of neutron physics more than 15 years ago. 

The author is grateful to all the participants in those discussions, the memory of which became an incentive for this work.

No funding sources have been involved in this research.

\end{document}